\documentstyle[12pt]{article}
\newcommand{\be}{\begin{equation}}
\newcommand{\ee}{\end{equation}}
\newcommand{\bea}{\begin{eqnarray}}
\newcommand{\eea}{\end{eqnarray}}
\newcommand{\sptwo}{1.4}
\newcommand{\doublespace}{\edef\baselinestretch{\sptwo}\Large\normalsize}
\newcommand{\newsection}[1]{
\section{#1}
\setcounter{equation}{0}}

\renewcommand{\theequation}{\thesection.\arabic{equation}}
\newcounter{newapp}
\renewcommand{\thenewapp}{\Alph{newapp}}
\begin{document}
\begin{center}{\Large\bf Superconformal Symmetry, The Supercurrent \\
And Non-BPS Brane Dynamics}
\end{center}
~\\
\begin{center}
T.E. Clark\footnote{e-mail address: clark@physics.purdue.edu}\\
{\it Department of Physics\\
Purdue University\\
West Lafayette, IN 47907-1396}\\
~\\
Muneto Nitta\footnote{e-mail address: nitta@th.phys.titech.ac.jp}\\
{\it Department of Physics\\
Tokyo Institute of Technology\\
Tokyo 152-8551, Japan}\\
~\\
T. ter Veldhuis\footnote{e-mail address: terveldhuis@macalester.edu}\\
{\it Department of Physics \& Astronomy\\
Macalester College\\
Saint Paul, MN 55105-1899}
\end{center}
\begin{center}
{\bf Abstract}
\end{center}
The Noether currents associated with the non-linearly realized super-Poincar\'{e} symmetries of the Green-Schwarz (Nambu-Goto-Akulov-Volkov) action for a non-BPS p=2 brane embedded in a N=1, D=4 target superspace are constructed.  The $R$ symmetry current, the supersymmetry currents, the energy-momentum tensor and the scalar central charge current are shown to be components of a world volume supercurrent.  The centrally extended superconformal transformations are realized on the Nambu-Goldstone boson and fermion fields of the non-BPS brane.  The superconformal currents form supersymmetry multiplets with the world volume conformal central charge current and special conformal current being the primary components of the supersymmetry multiplets containing all the currents.  Correspondingly the superconformal symmetry breaking terms form supersymmetry multiplets the components of which are obtainable as supersymmetry transformations of the primary currents' symmetry breaking terms.
\pagebreak
\doublespace

\newsection{Introduction}

The formation of a membrane in target space spontaneously breaks its symmetries to the isometries of the world volume and its complement.  The normal mode oscillations of the brane into the co-volume are described by world volume localized Nambu-Goldstone fields.  In the case of a target superspace, Goldstino modes must accompany the broken space translational Nambu-Goldstone boson modes.  The Green-Schwarz action \cite{GS} describes the dynamics of these world volume fields.   
In reference \cite{Clark:2002bh} the Nambu-Goto-Akulov-Volkov action for a non-BPS p=2 brane embedded in N=1, D=4 superspace was constructed via the nonlinear realization \cite{Volkov:jx}\cite{Coleman:sm}\cite{Ivanov:1999fw}
of the spontaneously broken super-Poincar\'e symmetries of the target superspace.  In reference \cite{Clark:2004kz} the Green-Schwarz action for this brane \cite{Sen:1999md}
was shown to be equivalent to the Nambu-Goto-Akulov-Volkov action by means of explicit nonlinear field redefinitions. The action described the motion of the brane in N=1, D=4 superspace through the brane localized Nambu-Goldstone boson field $\phi$ associated with motions in space directions transverse to the brane, hence in the direction 
of the broken space translation symmetry.  It also involved brane localized D=3 Majorana Goldstino 
fields $\theta_i$ and $\lambda_i$, $i=1,2$, describing brane oscillations in Grassmann directions of 
superspace which are associated with the completely broken N=1, D=4 
supersymmetry (SUSY).  The action, after application of the \lq\lq inverse Higgs 
mechanism" \cite{Ivanov:1975zq}, is the N=1, D=4 super-Poincar\'e invariant synthesis of 
the Akulov-Volkov \cite{Volkov:jx} and Nambu-Goto \cite{Nambu:1974zg} actions
\bea
\Gamma &=& -\sigma \int d^3 x \det{\hat{e}}\det{N} = -\sigma \int d^3 x \det{\hat{e}} \sqrt{ 1- \hat\nabla_a \phi \eta^{ab} \hat\nabla_b \phi} \cr
 &=& -\sigma \int d^3 x \det{\hat{e}} \sqrt{ 1- \lbrack \hat{e}_a^{-
1m}(\partial_m \phi + \bar\theta {\stackrel{\leftrightarrow}{\partial}}_m 
\lambda )\rbrack^2} ,
\label{CNtVAVNG}
\eea
where $\sigma$ is the brane tension and, in the static gauge, the world-volume coordinates are $x^m$, $m=0,1,2$.  The Akulov-Volkov dreibein is 
$\hat{e}_m^{~a} = \delta_m^{~a} + i\bar\theta \gamma^a \partial_m \theta 
+i\bar\lambda \gamma^a \partial_m \lambda .$  The Nambu-Goto dreibein is given by 
$N_a^{~b} = \delta_a^{~b} +\frac{\hat\nabla_a \phi \hat\nabla^b \phi}{(\hat\nabla \phi)^2} \left(\sqrt{1- (\hat\nabla \phi)^2} -1\right),$ in which the Nambu-Goldstone boson covariant derivative, $\hat\nabla_a \phi$, is defined as
\be
\hat\nabla_a \phi = \hat{\cal D}_a \phi + \bar\theta \hat{\cal D}_a \lambda - \hat{\cal D}_a \bar\theta \lambda 
= \hat{e}_a^{-1m} \left(\partial_m \phi + \bar\theta \stackrel{\leftrightarrow}{\partial}_m \lambda \right).
\ee
The partial covariant derivative, $\hat{\cal D}_a$, is defined by $\hat{\cal D}_a = \hat{e}_a^{-1m} \partial_m.$  The notation of reference \cite{Clark:2002bh} is followed throughout this paper. 

The purpose of this paper is to determine the symmetry currents of this action, which are those associated with the nonlinearly realized N=1, D=4 super-Poincar\'e symmetries.  Further, in accordance with the equivalent N=2, D=3 centrally extended SUSY algebra, it is shown that the $R$ symmetry current, the supersymmetry currents, the energy-momentum tensor and the scalar central charge current are components of a supercurrent.  That is a SUSY multiplet of currents.  The primary current in this multiplet is the $R$ current, the remaining derived currents can be obtained from it by SUSY transformations.  Hence, since the $R$ current is conserved, the conservation of the derived component SUSY currents, energy-momentum tensor and the central charge current is guaranteed.  Besides the spontaneously broken super-Poincar\'e currents, the centrally extended superconformal currents are constructed.  These also form supersymmetry multiplets.  Since the scale symmetries are explicitly broken, the superconformal current non-conservation terms are similarly related by the SUSY multiplet structure of the currents.  It is further shown that all superconformal as well as super-Poincar\'{e} currents are obtained as SUSY variations of the primary D=3 conformal central charge current and the primary D=3 special conformal current (both formerly comprise the D=4 special conformal symmetry current).

In section 2, Noether's theorem is stated and the Noether currents along with the variations of the Lagrangian (\ref{CNtVAVNG})
for the nonlinearly realized N=1, D=4 super-Poincar\'e symmetries are obtained.  Since the unbroken symmetries are those of the D=3 Poincar\'e group, it is useful to express the D=4 charges in terms of their D=3 Lorentz group transformation properties. 
Appendix A summarizes the N=1, D=4 superconformal algebra expressed as the corresponding centrally extended N=2, D=3 superconformal algebra.  In Appendix B the derivation of the nonlinear realization of this algebra on the world volume Nambu-Goldstone boson field, $\phi$, and Goldstino fields, $\theta$ and $\lambda$, is given for the case that the N=1, D=4 superconformal SU$(2,2|1)$ group is spontaneously broken to the D=3 SO(3,2) conformal group and U(1) $R$ symmetry (for fixed D=4 realizatons see \cite{sconformal}\cite{Clark:2004xj}).  The underlying short distance models \cite{Chibisov:1997rc}\cite{Eto:2004zc}
\cite{Isozumi:2004va} that give rise to the non-BPS domain wall formation should also explicitly break the scale symmetries both radiatively and by the dimensionful brane tension parameter in the models.  Hence physically no additional conformal Nambu-Goldstone modes occur.  However, the superconformal algebra, as discussed in Appendix B, requires the conformal central charge symmetry (the spatial component normal to the brane of the D=4 special conformal transformations) and the conformal SUSY transformations to have a spontaneously broken component.  Derivatives of the brane oscillation Nambu-Goldstone fields provide the required ground state expectation values for these broken symmetries and no additional fields are required in order to nonlinearly realize the above spontaneously broken symmetry transformations.  In section 3, Noether's theorem is again utilized in order to construct the remainder of the superconformal Noether currents.  Consistent with the charge algebra, these currents form SUSY multiplets of currents.  Since the superconformal currents are not conserved their explicit symmetry breaking terms also form SUSY multiplets.  As in the case of all superconformal currents, all superconformal symmetry breaking terms are shown to be given by SUSY transformations of the D=3 conformal central charge current non-conservation terms and the D=3 special conformal symmetry current non-conservation terms, the two independent primary component currents of the supermultiplets containing all currents.  Finally, four tables are included at the end of section 3 which summarize the symmetry charges and Noether currents associated with the Nambu-Goto-Akulov-Volkov action.  Also the SUSY variations of the currents and their (non-)conservation equations are recapitulated.
\pagebreak

\newsection{SUSY and the Supercurrent}

Noether's theorem provides a relation between the divergence of a symmetry current and the associated variation of the Lagrangian for a given transformation of the fields.  For a (Lie derivative) intrinsic symmetry transformation of the fields, denoted by $\delta \phi$, $\delta\theta$ and $\delta\lambda$ and such that $\delta (\partial_m \varphi^I) = \partial_m(\delta\varphi^I)$ where $\varphi^I$ stands for any of the fields $\varphi=(\phi, \theta , \lambda)$, the variation of the Lagrangian yields
\be
\delta {\cal L} = \partial_m j^m + w \Gamma .
\ee
The symmetry current $j^m=\sum_I \frac{\partial {\cal L}}{\partial\partial_m \varphi^I} \delta\varphi^I$ and the local Ward-Takahashi functional differential operator for the symmetry transformation (often referred to as Euler-Lagrange terms when acting on the action) is $w= \sum_I \delta\varphi^I \frac{\delta}{\delta\varphi^I}$ with $\Gamma$ the action, in this case, equation (\ref{CNtVAVNG}).  The intrinsic variation, $\delta \varphi = \varphi^\prime (x) - \varphi (x)$, is related to the total variation $\Delta \varphi = \varphi^\prime (x^\prime) - \varphi (x)$, by the subtraction of the first Taylor expansion term for the space-time coordinate variation; for the Lagrangian, this yields $\delta {\cal L} = \Delta {\cal L} - \delta x^m \partial_m {\cal L}.$  Using the chain rule for the Taylor term, the final form of Noether's theorem is obtained
\be
\partial_m J^m = \left[ \Delta {\cal L} + (\partial_m \delta x^m){\cal L}\right] -w\Gamma ,
\ee
where the Noether current, $J^m$, is defined as
\be
J^m \equiv j^m +\delta x^m {\cal L} = \sum_I \frac{\partial {\cal L}}{\partial\partial_m \varphi^I} \delta\varphi^I +\delta x^m {\cal L} .
\ee
Substituting the Lagrangian derivatives associated with equation (\ref{CNtVAVNG}), the Noether current is obtained
\be
J^m = -i\left[\delta\bar\theta \gamma^a \theta + \delta \bar\lambda \gamma^a \lambda\right]\left(\hat{e}_a^{-1m} 
+\frac{\hat\nabla_a\phi \hat\nabla^b \phi \hat{e}_b^{-1m}}{(\det{N})^2}\right){\cal L} -\frac{{\cal L}\hat\nabla^a \phi \hat{e}_a^{-1m}}{(\det{N})^2}\left[ \delta\phi -\delta\bar\theta\lambda +\delta\bar\lambda \theta\right] .
\label{Ncurrent}
\ee

The non-BPS p=2 brane action is world-volume space-time translationally invariant: $\Delta^p (a) {\cal L}= 0$.  Recalling the space-time translations of the fields from Appendix B, $\delta^p (a) x^m = a^m$ and $\delta^p (a) \varphi = -a^m\partial_m \varphi$, the Noether energy-momentum tensor is secured
\be
T^m_{~~a} = -{\cal L}\hat{e}_a^{-1m} -\frac{{\cal L}}{(\det{N})^2}\hat\nabla^b \phi \hat{e}_b^{-1m}  \hat\nabla_a\phi .
\ee
Conservation of brane energy-momentum yields $\partial_m T^m_{~~a} = w^p_a (x) \Gamma$, with the space-time translation 
Ward identity operator $w^p_a (x)= -\sum_I \partial_a \varphi^I (x) \frac{\delta}{\delta \varphi^I (x)}$.  (Integration over the world-volume leads to the global space-time translation Ward identity for the (tree level) one-particle irreducible generating functional, $\Gamma$, that is the effective action.)  Likewise, the spontaneously broken D=4 space translation symmetry normal to the brane, now expressed as the D=3 spontaneously broken scalar central charge, $Z$, symmetry, is conserved.  Its Noether current divergence is given simply by the $\phi$ field equation, hence
\be
Z^m = -\frac{{\cal L}}{(\det{N})^2}\hat\nabla^a \phi \hat{e}_a^{-1m} .
\ee
Since only $\phi$ transforms under $Z$, $\Delta^Z (z) \phi =z$, Noether's theorem yields $\partial_m Z^m =-w^Z (x) \Gamma$ with
$w^Z (x) = \frac{\delta}{\delta \phi (x)}$.  

All other currents, including those of the superconformal symmetries, can be expessed in terms of the D=3 energy-momentum tensor and the D=3 central charge symmetry current.  The general form of the Noether current, equation (\ref{Ncurrent}), becomes
\be
J^m = -T^m_{~~a} \left[ \delta x^a -i (\Delta \bar\theta \gamma^a \theta + \Delta \bar\lambda \gamma^a \lambda)\right] +
Z^m \left[ \Delta \phi - \Delta \bar\theta \lambda + \bar\theta \Delta\lambda\right].
\label{NcurrentTZ}
\ee
Similarly, using these currents, the variation of the Lagrangian takes on a simplified form
\be
\delta {\cal L} = -T^m_{~~a} \delta [\hat{e}_m^{~a}] + Z^m \delta [\hat{e}_m^{~a} \hat\nabla_a \phi ].
\label{deltaL}
\ee
Applying this to the case of space-time translations, for instance, yields the simple derivative formula for the Lagrangian
\be
\partial_n {\cal L} = -T^m_{~~a} \partial_n [\hat{e}_m^{~a}] + Z^m \partial_n [\hat{e}_m^{~a} \hat\nabla_a \phi ].
\ee

The brane Lagrangian is manifestly $R$ invariant resulting in the $R$ symmetry Noether current
\be
R^m = -2 T^m_{~~a} (\bar\theta \gamma^a \lambda) +i Z^m (\bar\theta \theta + \bar\lambda \lambda),
\ee
with the conservation equation $\partial_m R^m = -w^R (x) \Gamma$, where $w^R (x) = -i[\lambda (x) \frac{\delta}{\delta \theta (x)} -\theta (x)\frac{\delta}{\delta \lambda (x)}].$  Introducing $q-$ and $s-$ SUSY transformation parameters $\xi$ and $\eta$, respectively, that are Grassmann D=3 Majorana spinors, the SUSY variations become
\bea
\delta^{q,s} (\xi, \eta) x^m &=& -i(\bar\xi \gamma^m \theta + \bar\eta \gamma^m \lambda ) \equiv - \Lambda^m (\xi, \eta) \cr
\Delta^{q,s} (\xi, \eta) \theta_i &=& \xi_i \cr
\Delta^{q,s} (\xi, \eta) \lambda_i &=& \eta_i \cr
\Delta^{q,s} (\xi, \eta) \phi &=& -(\bar\xi \lambda - \bar\eta \theta) .
\eea
(Recall that $\delta^{q,s} (\xi, \eta) = \bar\xi_i \delta^q_i + \bar\eta_i \delta^s_i$, and so on.)  As usual for SUSY the Lagrangian has a total derivative intrinsic variation, $\delta^{q,s} (\xi, \eta) {\cal L}= \partial_m (\Lambda^m (\xi, \eta){\cal L}).$  The SUSY Noether currents are
\bea
Q^m (\xi) &=& \bar\xi_i Q^m_i (x) = +2i T^m_{~~a} (\bar\xi \gamma^a \theta ) -2 Z^m (\bar\xi \lambda) \cr
S^m (\eta) &=& \bar\eta_i S^m_i (x) = +2i T^m_{~~a} (\bar\eta \gamma^a \lambda ) +2 Z^m (\bar\eta \theta) .
\eea
Noether's theorem yields the SUSY Ward identities 
\be
\partial_m \left[ Q^m (\xi) + S^m (\eta)\right] = - w^{q,s} (x) \Gamma ,
\ee
with the SUSY Ward identity operator given as $w^{q,s} (x)=\sum_I \delta^{q,s} (\xi, \eta) \varphi^I (x) \frac{\delta}{\delta \varphi^I (x)}$.

The remaining super-Poincar\'{e} transformations are the unbroken world-volume Lorentz transformations with angular momentum tensor $M^m_{~ab} = \epsilon_{abc} M^{mc}$ and the spontaneously broken D=4 Lorentz transformations with D=3 current $N^m_{~~a}$.  The Lagrangian is invariant under D=3 Lorentz transformations yielding a conserved Noether current, $\partial_m M^m_{~ab} (x) = -w^M_{ab} (x) \Gamma$, with
\be
M^m_{~ab} = T^m_{~~a} x_b -T^m_{~~b}x_a + \frac{1}{2} T^m_{~~c} \left[ \bar\theta \gamma_{ab} \gamma^c \theta + \bar\lambda \gamma_{ab} \gamma^c \lambda \right] +i Z^m (\bar\theta \gamma_{ab} \lambda).
\ee
Under spontaneously broken D=4 Lorentz transformations the Lagrangian transforms into a total divergence $\delta^N (b) {\cal L} = \partial_m (b^m \phi {\cal L})$ resulting in a conserved Noether current $\partial_m N^m_{~~a} (x) = -w^n_a (x) \Gamma$, with
\be
N^m_{~~a} = T^m_{~~a} \phi +i T^m_{~~b}(\bar\theta \gamma_a^{~b} \lambda) -Z^m x_a .
\ee

For models in N=1, D=4 superspace with linearly represented SUSY transformations, the $R$ symmetry current, the SUSY currents and the energy-momentum tensor are components of a supercurrent superfield \cite{FZ}.  In addition all other superconformal currents are capable of being written in terms of space-time moments and SUSY covariant derivatives of this supercurrent.  The explicit and anomalous breaking terms of the $R$ and conformal symmetries also follow from space-time moments and SUSY covariant derivatives of the generalized trace of the supercurrent \cite{Clark:1978jx}\cite{PS1}\cite{Clark:1995bg}.  (For an analysis of the superconformal charges and the supercharge in supersymmetric quantum mechanics see reference \cite{Clark:2001zv}.)  In models with nonlinearly realized SUSY in D=4, the $R$ symmetry current, the SUSY currents and the energy-momentum tensor still form a SUSY multiplet \cite{Clark:1988es} with the superconformal currents being given by space-time moments and field monomials times the components of this supercurrent (explicitly the $R$ current and energy-momentum tensor)\cite{Clark:2004xj}.  As shown below, in brane models these currents along with the central charge symmetry current form a SUSY multiplet.  As well, all superconformal currents are shown to be world volume coordinate moments and field monomials times the components of the supercurrent (explicitly the energy-momentum tensor and the central charge current as in equation (\ref{NcurrentTZ})).  In general starting with a current $J^m$, its $q-$ and $s-$ SUSY transformations yield a relation of the form
\bea
\delta^{q,s} (\xi, \eta) J^m &=& \partial_n \left[ J^m \Lambda^n (\xi, \eta) - J^n \Lambda^m (\xi, \eta)\right] +\Lambda^m (\xi, \eta) \partial_n J^n \cr
 & & \qquad\qquad\qquad\qquad + {\cal J}^m (\xi, \eta), 
\label{currentmultiplet}
\eea
where ${\cal J}^m (\xi, \eta)$ is another of the super-Poincar\'{e} or superconformal currents or zero.  If zero, then that $J^m$ is the last (or highest weight) component in a supermultiplet.  The first set of terms on the right hand side are current improvement terms since they are algebraically divergenceless.  They can be added to ${\cal J}^m (\xi, \eta)$ to define an improved and still conserved current
\be
{\cal J}^m_{\rm Imp} (\xi, \eta) = {\cal J}^m (\xi, \eta) +\partial_n \left[ J^m \Lambda^n (\xi, \eta) - J^n \Lambda^m (\xi, \eta)\right] ,
\ee
and $\partial_m {\cal J}^m_{\rm Imp} (\xi, \eta) =\partial_m {\cal J}^m (\xi, \eta).$  The improvement terms will be kept explicit in what follows.  The set of currents which begins with some primary current $J^m$ and continues upon SUSY variation to yield other currents until the last component current is reached comprise a supermultiplet of currents.  Every superconformal current will form such a multiplet, even if the primary current is the sole member.  Since differentiation commutes with intrinsic variation, the (non-) conservation of the primary current implies the (possible non-)conservation of the higher component currents.  The second term on the right hand side of equation (\ref{currentmultiplet}) along with the relation between intrinsic and total variation results in the higher component conservation equation
\be
\partial_m {\cal J}^m (\xi, \eta) = \Delta^{q,s} (\xi, \eta) (\partial_m J^m) + (\partial_n\delta^{q,s} (\xi, \eta) x^n)(\partial_m J^m).
\label{Jbreaking}
\ee
So the currents in any multiplet can be obtained from the primary current by SUSY variation.  Likewise, the breaking of any symmetry in a multiplet can be obtained from the primary current symmetry breaking by SUSY variation.

The $R^m$, $Q^m_i$, $S^m_i$, $T^m_{~~a}$ and $Z^m$ currents form the supercurrent multiplet with the $R$ current as the primary component.  Since $R^m$ is conserved, so too are the other currents in this supercurrent as verified from their explicit construction.  The SUSY variations are found to be (using $\psi_i$ as a Grassmann, D=3 Majorana spinor parameter)
\bea
\delta^{q,s} (\xi, \eta) R^m &=&  \partial_n \left[ R^m \Lambda^n (\xi, \eta) - R^n \Lambda^m (\xi, \eta)\right] +\Lambda^m (\xi, \eta) \partial_n R^n \cr
 & & \qquad\qquad\qquad -iQ^m (\eta) +i S^m (\xi) \cr
\delta^{q,s} (\xi, \eta) Q^m (\psi) &=&  \partial_n \left[ Q^m (\psi) \Lambda^n (\xi, \eta) - Q^n (\psi) \Lambda^m (\xi, \eta)\right] +\Lambda^m (\xi, \eta) \partial_n Q^n (\psi) \cr
 & & \qquad\qquad\qquad +2i T^m_{~~a} (\bar\psi \gamma^a \xi) -2 Z^m (\bar\psi \eta) \cr
\delta^{q,s} (\xi, \eta) S^m (\psi) &=&  \partial_n \left[ S^m (\psi) \Lambda^n (\xi, \eta) - S^n (\psi) \Lambda^m (\xi, \eta)\right] +\Lambda^m (\xi, \eta) \partial_n S^n (\psi) \cr
 & & \qquad\qquad\qquad +2i T^m_{~~a} (\bar\psi \gamma^a \eta) +2 Z^m (\bar\psi \xi) \cr
\delta^{q,s} (\xi, \eta) T^m_{~~a} &=&  \partial_n \left[ T^m_{~~a} \Lambda^n (\xi, \eta) - T^n_{~~a} \Lambda^m (\xi, \eta)\right] +\Lambda^m (\xi, \eta) \partial_n T^n_{~~a} \cr
\delta^{q,s} (\xi, \eta) Z^m &=&  \partial_n \left[ Z^m \Lambda^n (\xi, \eta) - Z^n \Lambda^m (\xi, \eta)\right] +\Lambda^m (\xi, \eta) \partial_n Z^n .
\eea
From the super-Poincar\'{e} symmetry group perspective, the angular momentum tensor and the broken Lorentz transformation current are primary currents whose SUSY multiplets involve the SUSY currents, the energy-momentum tensor and the central charge current
\bea
\delta^{q,s} (\xi, \eta) M^m_{ab} &=&  \partial_n \left[ M^m_{ab} \Lambda^n (\xi, \eta) - M^n_{ab} \Lambda^m (\xi, \eta)\right] +\Lambda^m (\xi, \eta) \partial_n M^n_{ab} \cr
 & & \qquad\qquad\qquad -i(\bar\xi \gamma_{ab}Q^m) +i (\bar\eta \gamma_{ab}S^m) \cr
\delta^{q,s} (\xi, \eta) N^m_{~~a} &=&  \partial_n \left[ N^m_{~~a} \Lambda^n (\xi, \eta) - N^n_{~~a} \Lambda^m (\xi, \eta)\right] +\Lambda^m (\xi, \eta) \partial_n N^n_{~~a} \cr
 & & \qquad\qquad\qquad -i(\bar\xi \gamma_a S^m)  +i (\bar\eta \gamma_a Q^m) .
\eea
\pagebreak

\newsection{Superconformal Currents and SUSY Multiplets}

Noether's theorem applies to the construction of the superconformal currents and their (non-)conservation equations as well.  
Utilizing equation (\ref{deltaL}), the Lagrangian varies under scale transformations as $\delta^D (\epsilon) {\cal L} = -\epsilon x^m \partial_m {\cal L}=3\epsilon {\cal L} -\partial_m (\epsilon x^m {\cal L})$.  This results in a dilatation Noether current
\be
D^m = -T^m_{~~a} x^a +Z^m \phi .
\ee
The scale symmetry is explicitly violated so that the divergence of the dilatation current obeys the broken Ward identity
\bea
\partial_m D^m &=& -T^m_{~~a} \hat{e}_m^{~a} + Z^m \hat{e}_m^{~a} \hat\nabla_a \phi - w^D \Gamma \cr
 &=& 3 \sigma \frac{\partial}{\partial \sigma} {\cal L} -w^D \Gamma ,
\eea
where as usual the local Ward identity dilatation operator is $w^D (x)= \sum_I \delta^D \varphi^I (x) \frac{\delta}{\delta \varphi^I}$.  

The D=4 special conformal transformations lead to the D=3 special conformal transformations and the conformal central charge symmetry.  The Noether current for the latter is secured as
\be
Y^m = +2 T^m_{~~a} \left[ x^a \phi +i \epsilon^{abc} x_b (\bar\theta \gamma_c \lambda)\right] -Z^m \left[ x^2 +\phi^2 - (\bar\theta \theta)(\bar\lambda \lambda)\right] .
\ee
Noether's theorem yields the explicitly broken (but realized as spontaneously broken, as discussed in Appendix B) $Y$ symmetry Ward identity
\be
\partial_m Y^m = \left[ \Delta^Y {\cal L} +(\partial_m \delta^Y x^m){\cal L}\right] - w^Y \Gamma ,
\ee\
where $w^Y = \sum_I \delta^Y \varphi^I \frac{\delta}{\delta \varphi^I}$.  The breaking terms are obtained from equation (\ref{deltaL})
\bea
\delta^Y {\cal L} &=& 2 x^m \partial_m (\phi {\cal L}) +T^m_{~~a} \left[ 2x^a (\hat{e}_m^{~b} \hat\nabla_b \phi) -2i \epsilon_m^{~~ab}(\bar\theta \gamma_b \lambda)\right. \cr
 & &\left. \qquad\qquad\qquad\qquad +2i(\bar\theta \theta - \bar\lambda \lambda)(\bar\theta \gamma^a \stackrel{\leftrightarrow}{\partial}_m \lambda)\right] -2 Z^m \hat{e}_m^{~a} x_a .
\eea
Applying this to the divergence equation, the non-conservation of the $Y$ current is found
\pagebreak
\bea
\left[ \Delta^Y {\cal L} +(\partial_m \delta^Y x^m){\cal L}\right] &=& -6\phi {\cal L} +T^m_{~~a} \left[ 2x^a (\hat{e}_m^{~b} \hat\nabla_b \phi) -2i \epsilon_m^{~~ab}(\bar\theta \gamma_b \lambda)\right. \cr
 & &\left. \qquad\qquad\qquad +2i(\bar\theta \theta - \bar\lambda \lambda)(\bar\theta \gamma^a \stackrel{\leftrightarrow}{\partial}_m \lambda)\right] -2 Z^m \hat{e}_m^{~a} x_a .\cr
 & & 
\label{brokenY}
\eea
The special conformal current $K^m_{~~a}$ is determined to be
\bea
K^m_{~~a} &=& -T^m_{~~b} \left[ 2 x^b x_a - x^2 \delta^b_{~a} + \phi^2 \delta^b_{~a} + (\bar\theta \theta - \bar\lambda \lambda) 
\delta^b_{~a} \right. \cr
 & &\left. \qquad\qquad\qquad - 2i \epsilon_a^{~bc} \phi (\bar\theta \gamma_c \lambda) - \epsilon_a^{~bc} x_c (\bar\theta \theta - \bar\lambda \lambda) \right] \cr
 & &\qquad\qquad\qquad\qquad\qquad +Z^m \left[ 2x_a \phi -2i \epsilon_a^{~bc} x_b (\bar\theta \gamma _c \lambda)\right] .
\eea
The special conformal current is explicitly not conserved, obeying the divergence equation
\be
\partial_m K^m_{~~a} = \left[ \Delta^k_{a} {\cal L} + (\partial_m \delta^k_{a} x^m) {\cal L}\right] -w^k_{a} \Gamma ,
\ee
where the Ward identity term $w^k_a (x) = \sum_I \delta^k_a \varphi^I \frac{\delta}{\delta \varphi^I}$.
The special conformal variation, with transformation parameter $\epsilon^a$, of the Lagrangian is determined directly using the field variations found in Appendix B.  This yields the total variation of the Lagrangian, $\Delta^k (\epsilon) {\cal L} = \delta^k (\epsilon) {\cal L} + (\delta^k (\epsilon) x^m )\partial_m {\cal L}$, and from this the current breaking terms are found
\bea
\left[ \Delta^k (\epsilon) {\cal L} + (\partial_m \delta^k (\epsilon) x^m) {\cal L}\right]&=&6\epsilon^m x_m {\cal L} \cr
 & &-T_{~~a}^m \left[ 2\epsilon^a \phi [\hat{e}_m^{~b} \hat\nabla_b \phi ] + 2(\epsilon_b x^a - \epsilon^a x_b) \hat{e}_m^{~b}    -\epsilon^a \partial_m [(\bar\theta\theta) (\bar\lambda\lambda)] \right.\cr
 & &\left. \qquad\qquad +(\bar\theta\theta + \bar\lambda\lambda)\epsilon^{a}_{~bc} \epsilon^c [ \hat{e}_m^{~b} -2\delta^b_{~m}] +4(\bar\theta \gamma^a \lambda)\partial_m (\bar\theta \rlap{/}{\epsilon} \lambda) \right.\cr
 & &\left. \qquad\qquad\qquad\qquad - 2i \epsilon^{abc} \epsilon_c (\bar\theta \gamma_b \lambda) [\hat{e}_m^{~d}\hat\nabla_d \phi]\right] \cr
 & & + Z^m \left[2 \epsilon_a \phi \hat{e}_m^{~a} - 2i \epsilon_{mab} \epsilon^b (\bar\theta \gamma^a \lambda)
-2i (\bar\theta\theta - \bar\lambda\lambda ) [\bar\theta \rlap{/}{\epsilon} \stackrel{\leftrightarrow}{\partial}_m \lambda] \right] .\cr
 & & 
\label{brokenk}
\eea

Finally the $u_i-$ and $v_i-$ conformal SUSY spinor currents are obtained using equation (\ref{NcurrentTZ}) (with the explicitly broken conformal SUSY realized as a spontaneously broken symmetry, as discussed in Appendix B)
\bea
U^m_i &=& 2 T^m_{~~a} \left[ x_b (\gamma^b \gamma^a \theta)_i -i\phi(\gamma^a \lambda)_i +i (\bar\lambda \lambda) (\gamma^a \theta)_i\right] \cr
 & &\qquad\qquad\qquad\qquad +2 Z^m \left[ i(\rlap{/}{x}\lambda)_i -\phi \theta_i + (\bar\theta \theta) \lambda_i \right] \cr
V^m_i &=& -2 T^m_{~~a} \left[ x_b (\gamma^b \gamma^a \lambda)_i +i\phi(\gamma^a \theta)_i +i (\bar\theta \theta) (\gamma^a \lambda)_i\right] \cr
 & &\qquad\qquad\qquad\qquad +2 Z^m \left[ i(\rlap{/}{x}\theta)_i +\phi \lambda_i + (\bar\lambda \lambda) \theta_i \right] .
\eea
The non-conservation equations have the form
\bea
\partial_m U^m_i &=& \left[\Delta^u_i {\cal L} + (\partial_m \delta^u_i x^m ){\cal L}\right] - w^u_i \Gamma \cr
\partial_m V^m_i &=& \left[\Delta^v_i {\cal L} + (\partial_m \delta^v_i x^m ){\cal L}\right] - w^v_i \Gamma ,
\eea
where the Ward identity operators are $w^{u,v}_i (x) = \sum_I \delta^{u,v}_i \varphi^I \frac{\delta}{\delta \varphi^I}$.
The superconformal symmetry breaking terms can be determined directly, however, as shown below, the $U^m_i$ and $V^m_i$ currents are components of a conformal current SUSY multiplet and so the breaking can be found from the SUSY variation of the primary current's breaking terms.

As in the case of the super-Poincar\'{e} symmetry currents, the superconformal currents belong to $q-$ and $s-$ SUSY multiplets consistent with the superconformal algebra.  The SUSY variations of the currents are just parts of the superconformal 
charge-current algebra.  In the superconformal case, the conformal central charge current, $Y^m$, and the special conformal current, $K^m_{~~a}$, act as independent primary currents from which the remainder of the superconformal and even the super-Poincar\'{e} currents can be obtained.  The $q-$ and $s-$ SUSY variations of the $Y^m$ and $K^m_{~~a}$ currents are 
\bea
\delta^{q,s} (\xi, \eta) Y^m &=& \partial_n \left[ Y^m \Lambda^n (\xi, \eta) - Y^n \Lambda^m (\xi, \eta)\right] +\Lambda^m (\xi, \eta) \partial_n Y^n \cr
 & & \qquad\qquad\qquad\qquad +(\bar\eta U^m) +(\bar\xi V^m) \cr
\delta^{q,s} (\xi, \eta) K^m_{~~a} &=& \partial_n \left[ K^m_{~~a} \Lambda^n (\xi, \eta) - K^n_{~~a} \Lambda^m (\xi, \eta)\right] +\Lambda^m (\xi, \eta) \partial_n K^n_{~~a} \cr
 & & \qquad\qquad\qquad\qquad + i (\bar\xi \gamma_a U^m) - i (\bar\eta \gamma_a V^m) .
\eea
The SUSY variations of the conformal SUSY currents lead to the dilatation current and the super-Poincar\'{e} Lorentz and $R$ currents, introducing the D=3 Majorana spinor parameter $\psi_i$, 
\bea
\delta^{q,s} (\xi, \eta) U^m (\psi) &=&  \partial_n \left[ U^m (\psi) \Lambda^n (\xi, \eta) - U^n (\psi) \Lambda^m (\xi, \eta)\right] +\Lambda^m (\xi, \eta) \partial_n U^n (\psi) \cr
 & & \qquad\qquad\qquad +2i  (\bar\eta\gamma^a \psi) N^m_{~~a} -3i (\bar\eta\psi) R^m  \cr
 & & \qquad\qquad\qquad\qquad +2i  (\bar\xi\gamma^a \psi) M^m_{~~a} -2 (\bar\xi\psi) D^m  \cr
\delta^{q,s} (\xi, \eta) V^m (\psi) &=&  \partial_n \left[ V^m (\psi) \Lambda^n (\xi, \eta) - V^n (\psi) \Lambda^m (\xi, \eta)\right] +\Lambda^m (\xi, \eta) \partial_n V^n (\psi) \cr
 & & \qquad\qquad\qquad +2i  (\bar\xi\gamma^a \psi) N^m_{~~a} -3i (\bar\xi\psi) R^m  \cr
 & & \qquad\qquad\qquad\qquad -2i  (\bar\eta\gamma^a \psi) M^m_{~~a} +2 (\bar\eta\psi) D^m  .
\eea
Completing the multiplet, the SUSY variation of the dilatation current yields the $q-$ and $s-$ SUSY currents
\bea
\delta^{q,s} (\xi, \eta) D^m &=& \partial_n \left[ D^m \Lambda^n (\xi, \eta) - D^n \Lambda^m (\xi, \eta)\right] +\Lambda^m (\xi, \eta) \partial_n D^n \cr
 & & \qquad\qquad\qquad + \frac{1}{2} (\bar\xi Q^m) + \frac{1}{2} (\bar\eta S^m).
\eea
From the conformal central charge symmetry breaking terms, equation (\ref{brokenY}), or the special conformal symmetry breaking
terms, equation (\ref{brokenk}), the remainder of the superconformal symmetry breaking terms can be obtained by means of SUSY variations of the primary current's breaking terms as per equation (\ref{Jbreaking}).

The Noether currents associated with the non-linearly realized super-Poincar\'{e} symmetries of the Nambu-Goto-Akulov-Volkov action, equation (\ref{CNtVAVNG}), for a non-BPS p=2 brane embedded in a N=1, D=4 target superspace are summarized in Table 1.  The target space symmetry transformation generators are given with their D=3 Lorentz group decomposition and the related Noether currents are listed.  In Table 2 the forms of these Noether currents are specified for the effective action given in equation (\ref{CNtVAVNG}).  The SUSY variations of the Noether currents are listed in Table 3.  The $R$ symmetry current is the primary current for the supercurrent multiplet of centrally extended super-translation Noether currents.  The $R$ symmetry current, the $q$-SUSY and $s$-SUSY currents, the energy-momentum tensor and the scalar central charge current are the components of this supercurrent.  Because the $R$ current is conserved, the conservation of the derived component SUSY currents, energy-momentum tensor and central charge current is guaranteed.  This, as well as the (non-)conservation equations of the remaining super-Poincar\'e and centrally extended superconformal currents are listed in Table 4.  The superconformal currents also form supersymmetry multiplets.  Since the scale symmetries are explicitly broken, the superconformal current non-conservation terms are similarly related by the SUSY multiplet structure of the currents.  It was further shown that all superconformal, as well as super-Poincar\'{e}, currents are obtained as SUSY variations of the primary D=3 conformal central charge current and the primary D=3 special conformal current.

\begin{center}
{\bf Table 1: Symmetries, Charges and Currents}
\end{center}
{\hspace{-0.4in}
\begin{tabular}{| c | c | c |}
\hline {\bf D=4 Symmetry and Charge} & {\bf D=3 Symmetry and Charge} & {\bf World Volume} 
\raisebox{-1ex}{\rule{0cm}{6ex}}\\
& & {\bf Noether Current}
\raisebox{-2ex}{\rule{0cm}{3ex}}\\
\hline\hline Target Space Translations & World Volume Translations $p_a = P_{\mu = a}$ & $T^{m}_{~~a}$ 
\raisebox{-2ex}{\rule{0cm}{5ex}}\\
\cline{2-3}$P_\mu$ & Central Charge $Z= P_{\mu=3}$ & $Z^m$ 
\raisebox{-2ex}{\rule{0cm}{5ex}}\\
\hline Lorentz Transformations & World Volume Lorentz  &  
\raisebox{-2ex}{\rule{0cm}{5ex}}\\
 & Transformations $M_{ab}= M_{\mu=a~\nu=b}$ & $M^m_{~ab}$
\raisebox{-2ex}{\rule{0cm}{3ex}}\\
\cline{2-3}$M_{\mu\nu}$ & Broken Lorentz  & 
\raisebox{-2ex}{\rule{0cm}{5ex}}\\ 
 & Automorphisms $N_a =-M_{\mu=a~\nu=3}$ & $N^m_{~~a}$ 
\raisebox{-2ex}{\rule{0cm}{3ex}}\\
\hline N=1, D=4 SUSY & N=2, D=3 SUSY & $Q^m_i$ 
\raisebox{-2ex}{\rule{0cm}{5ex}}\\
\cline{3-3}$Q_\alpha$, $\bar{Q}_{\dot\alpha}$  & $\pmatrix{
q_i \cr
s_i \cr} = \frac{1}{2} e^{i\frac{\pi}{4}} \pmatrix{
\sigma & i\sigma\sigma_z \cr
-i\sigma & -\sigma\sigma_z \cr} \pmatrix{Q_\alpha \cr
\bar{Q}^{\dot\alpha} \cr} $ & $S^m_i$
\raisebox{-3ex}{\rule{0cm}{5ex}}\\
\hline $R$-Transformation $R$ & $R$-Transformation $R=R$ & $ R^m$ 
\raisebox{-2ex}{\rule{0cm}{5ex}}\\
\hline Dilatation $D$ & Dilatation $D = D$ & $D^m$ 
\raisebox{-2ex}{\rule{0cm}{5ex}}\\
\hline Special Conformal & Special Conformal $k_a =K_{\mu=a}$ & $K^m_{~~a}$ 
\raisebox{-2ex}{\rule{0cm}{5ex}}\\
\cline{2-3}$K_\mu$ & Conformal Central Charge $Y= K_{\mu=3}$ & $Y^m$ 
\raisebox{-2ex}{\rule{0cm}{5ex}}\\
\hline N=1, D=4 Superconformal & N=2, D=3 Superconformal  & $U^m_i$ 
\raisebox{-2ex}{\rule{0cm}{5ex}}\\
\cline{3-3}$S_\alpha$, $\bar{S}_{\dot\alpha}$  & $\pmatrix{
u_i \cr
v_i \cr} = \frac{1}{2} e^{i\frac{\pi}{4}} \pmatrix{
\sigma & i\sigma\sigma_z \cr
-i\sigma & -\sigma\sigma_z \cr} \pmatrix{S_\alpha \cr
\bar{S}^{\dot\alpha} \cr}$ & $V^m_i$ 
\raisebox{-3ex}{\rule{0cm}{5ex}}\\ 
\hline
\end{tabular}}
\newpage
\begin{center}
{\bf Table 2: Noether Currents}
\end{center}
{\hspace{-0.25in}
\begin{tabular}{| c | l |}
\hline {\bf Noether Current} & {\hspace{1.0in}\bf Form of Noether Current}    
\raisebox{-2ex}{\rule{0cm}{5ex}}\\
$J^m$ & $J^m = -T^m_{~~a} \left[ \delta x^a -i (\Delta \bar\theta \gamma^a \theta + \Delta \bar\lambda \gamma^a \lambda)\right] + Z^m \left[ \Delta \phi - \Delta \bar\theta \lambda + \bar\theta \Delta\lambda\right]$
\raisebox{-2ex}{\rule{0cm}{5ex}}\\
\hline\hline $T^{m}_{~~a}$ & $T^m_{~~a} = -{\cal L}\hat{e}_a^{-1m} -\frac{{\cal L}}{(\det{N})^2}\hat\nabla^b \phi \hat{e}_b^{-1m}  \hat\nabla_a\phi$ 
\raisebox{-2ex}{\rule{0cm}{5ex}}\\
\hline $Z^m$ & $Z^m = -\frac{{\cal L}}{(\det{N})^2}\hat\nabla^a \phi \hat{e}_a^{-1m}$ 
\raisebox{-2ex}{\rule{0cm}{5ex}}\\
\hline $M^m_{~ab}$ & $M^m_{~ab} = T^m_{~~a} x_b -T^m_{~~b}x_a + \frac{1}{2} T^m_{~~c} \left[ \bar\theta \gamma_{ab} \gamma^c \theta + \bar\lambda \gamma_{ab} \gamma^c \lambda \right] +i Z^m (\bar\theta \gamma_{ab} \lambda)$ 
\raisebox{-2ex}{\rule{0cm}{5ex}}\\
\hline $N^m_{~~a}$ & $N^m_{~~a} = T^m_{~~a} \phi +i T^m_{~~b}(\bar\theta \gamma_a^{~b} \lambda) -Z^m x_a $  
\raisebox{-2ex}{\rule{0cm}{5ex}}\\
\hline $Q^m_i$ & $Q^m (\xi) = \bar\xi_i Q^m_i (x) = +2i T^m_{~~a} (\bar\xi \gamma^a \theta ) -2 Z^m (\bar\xi \lambda)$ 
\raisebox{-2ex}{\rule{0cm}{5ex}}\\
\hline $S^m_i$ & $S^m (\eta) = \bar\eta_i S^m_i (x) = +2i T^m_{~~a} (\bar\eta \gamma^a \lambda ) +2 Z^m (\bar\eta \theta)$ 
\raisebox{-2ex}{\rule{0cm}{5ex}}\\
\hline $R^m$ & $R^m = -2 T^m_{~~a} (\bar\theta \gamma^a \lambda) +i Z^m (\bar\theta \theta + \bar\lambda \lambda)$ 
\raisebox{-2ex}{\rule{0cm}{5ex}}\\
\hline $D^m$ & $D^m = -T^m_{~~a} x^a +Z^m \phi$ 
\raisebox{-2ex}{\rule{0cm}{5ex}}\\
\hline $K^m_{~~a}$ & $K^m_{~~a} = -T^m_{~~b} \left[ 2 x^b x_a - x^2 \delta^b_{~a} + \phi^2 \delta^b_{~a} + (\bar\theta \theta - \bar\lambda \lambda) 
\delta^b_{~a}\right.$ 
\raisebox{-2ex}{\rule{0cm}{5ex}}\\
 & $\left. \qquad\qquad\qquad\qquad\qquad\qquad - 2i \epsilon_a^{~bc} \phi (\bar\theta \gamma_c \lambda) - \epsilon_a^{~bc} x_c (\bar\theta \theta - \bar\lambda \lambda) \right]$
\raisebox{-2ex}{\rule{0cm}{5ex}}\\
 & $\qquad\qquad +Z^m \left[ 2x_a \phi -2i \epsilon_a^{~bc} x_b (\bar\theta \gamma _c \lambda)\right]$
\raisebox{-2ex}{\rule{0cm}{5ex}}\\
\hline $Y^m$ & $Y^m = +2 T^m_{~~a} \left[ x^a \phi +i \epsilon^{abc} x_b (\bar\theta \gamma_c \lambda)\right] -Z^m \left[ x^2 +\phi^2 - (\bar\theta \theta)(\bar\lambda \lambda)\right]$ 
\raisebox{-2ex}{\rule{0cm}{5ex}}\\
\hline $U^m_i$ & $U^m_i = 2 T^m_{~~a} \left[ x_b (\gamma^b \gamma^a \theta)_i -i\phi(\gamma^a \lambda)_i +i (\bar\lambda \lambda) (\gamma^a \theta)_i\right] $
\raisebox{-2ex}{\rule{0cm}{5ex}}\\
 & $\qquad\qquad +2 Z^m \left[ i(\rlap{/}{x}\lambda)_i -\phi \theta_i + (\bar\theta \theta) \lambda_i \right]$
\raisebox{-2ex}{\rule{0cm}{5ex}}\\
\hline $V^m_i$ & $V^m_i = -2 T^m_{~~a} \left[ x_b (\gamma^b \gamma^a \lambda)_i +i\phi(\gamma^a \theta)_i +i (\bar\theta \theta) (\gamma^a \lambda)_i\right]$ 
\raisebox{-2ex}{\rule{0cm}{5ex}}\\
 & $\qquad\qquad +2 Z^m \left[ i(\rlap{/}{x}\theta)_i +\phi \lambda_i + (\bar\lambda \lambda) \theta_i \right]$ 
\raisebox{-2ex}{\rule{0cm}{5ex}}\\ 
\hline
\end{tabular}}
\newpage
\begin{center}
{\bf Table 3: SUSY Multiplets of Noether Currents}
\end{center}
{\hspace{-0.5in}
\begin{tabular}{| c | l |}
\hline {\bf Noether Current} & \hspace{1.0in}{\bf SUSY Transformation of Noether Current}    
\raisebox{-1ex}{\rule{0cm}{4ex}}\\
$J^m$ & $\delta^{q,s} (\xi, \eta) J^m = \partial_n \left[ J^m \Lambda^n (\xi, \eta) - J^n \Lambda^m (\xi, \eta)\right] +\Lambda^m (\xi, \eta) \partial_n J^n $
\raisebox{-1ex}{\rule{0cm}{4ex}}\\
 & \hspace{2.0in}$+ {\cal J}^m (\xi, \eta)$
\raisebox{-2ex}{\rule{0cm}{4ex}}\\
\hline\hline $T^{m}_{~~a}$ & $\delta^{q,s} (\xi, \eta) T^m_{~~a} =  \partial_n \left[ T^m_{~~a} \Lambda^n (\xi, \eta) - T^n_{~~a} \Lambda^m (\xi, \eta)\right] +\Lambda^m (\xi, \eta) \partial_n T^n_{~~a}$ 
\raisebox{-2ex}{\rule{0cm}{5ex}}\\
\hline $Z^m$ & $\delta^{q,s} (\xi, \eta) Z^m =  \partial_n \left[ Z^m \Lambda^n (\xi, \eta) - Z^n \Lambda^m (\xi, \eta)\right] +\Lambda^m (\xi, \eta) \partial_n Z^n$ 
\raisebox{-2ex}{\rule{0cm}{5ex}}\\
\hline $M^m_{~ab}$ & $\delta^{q,s} (\xi, \eta) M^m_{ab} =  \partial_n \left[ M^m_{ab} \Lambda^n (\xi, \eta) - M^n_{ab} \Lambda^m (\xi, \eta)\right] +\Lambda^m (\xi, \eta) \partial_n M^n_{ab} $ 
\raisebox{-1ex}{\rule{0cm}{4ex}}\\
 & \hspace{2.0in}$-i(\bar\xi \gamma_{ab}Q^m) +i (\bar\eta \gamma_{ab}S^m) $ 
\raisebox{-2ex}{\rule{0cm}{4ex}}\\
\hline $N^m_{~~a}$ & $\delta^{q,s} (\xi, \eta) N^m_{~~a} = \partial_n \left[ N^m_{~~a} \Lambda^n (\xi, \eta) - N^n_{~~a} \Lambda^m (\xi, \eta)\right] +\Lambda^m (\xi, \eta) \partial_n N^n_{~~a}$  
\raisebox{-1ex}{\rule{0cm}{4ex}}\\
 & \hspace{2.0in}$-i(\bar\xi \gamma_a S^m)  +i (\bar\eta \gamma_a Q^m)$ 
\raisebox{-2ex}{\rule{0cm}{4ex}}\\
\hline $Q^m_i$ & $\delta^{q,s} (\xi, \eta) Q^m (\psi) =  \partial_n \left[ Q^m (\psi) \Lambda^n (\xi, \eta) - Q^n (\psi) \Lambda^m (\xi, \eta)\right] +\Lambda^m (\xi, \eta) \partial_n Q^n (\psi) $ 
\raisebox{-1ex}{\rule{0cm}{4ex}}\\
 & \hspace{2.0in}$+2i T^m_{~~a} (\bar\psi \gamma^a \xi) -2 Z^m (\bar\psi \eta)$ 
\raisebox{-2ex}{\rule{0cm}{4ex}}\\
\hline $S^m_i$ & $\delta^{q,s} (\xi, \eta) S^m (\psi) = \partial_n \left[ S^m (\psi) \Lambda^n (\xi, \eta) - S^n (\psi) \Lambda^m (\xi, \eta)\right] +\Lambda^m (\xi, \eta) \partial_n S^n (\psi) $ 
\raisebox{-1ex}{\rule{0cm}{4ex}}\\
 & \hspace{2.0in}$+2i T^m_{~~a} (\bar\psi \gamma^a \eta) +2 Z^m (\bar\psi \xi)$ 
\raisebox{-2ex}{\rule{0cm}{4ex}}\\
\hline $R^m$ & $\delta^{q,s} (\xi, \eta) R^m =  \partial_n \left[ R^m \Lambda^n (\xi, \eta) - R^n \Lambda^m (\xi, \eta)\right] +\Lambda^m (\xi, \eta) \partial_n R^n $ 
\raisebox{-1ex}{\rule{0cm}{4ex}}\\
 & \hspace{2.0in}$-iQ^m (\eta) +i S^m (\xi) $ 
\raisebox{-2ex}{\rule{0cm}{4ex}}\\
\hline $D^m$ & $\delta^{q,s} (\xi, \eta) D^m = \partial_n \left[ D^m \Lambda^n (\xi, \eta) - D^n \Lambda^m (\xi, \eta)\right] +\Lambda^m (\xi, \eta) \partial_n D^n $ 
\raisebox{-1ex}{\rule{0cm}{4ex}}\\
 & \hspace{2.0in}$+ \frac{1}{2} (\bar\xi Q^m) + \frac{1}{2} (\bar\eta S^m)$ 
\raisebox{-2ex}{\rule{0cm}{4ex}}\\
\hline $K^m_{~~a}$ & $\delta^{q,s} (\xi, \eta) K^m_{~~a} = \partial_n \left[ K^m_{~~a} \Lambda^n (\xi, \eta) - K^n_{~~a} \Lambda^m (\xi, \eta)\right] +\Lambda^m (\xi, \eta) \partial_n K^n_{~~a} $ 
\raisebox{-1ex}{\rule{0cm}{4ex}}\\
 & \hspace{2.0in}$+ i (\bar\xi \gamma_a U^m) - i (\bar\eta \gamma_a V^m)$ 
\raisebox{-2ex}{\rule{0cm}{4ex}}\\
\hline $Y^m$ & $\delta^{q,s} (\xi, \eta) Y^m = \partial_n \left[ Y^m \Lambda^n (\xi, \eta) - Y^n \Lambda^m (\xi, \eta)\right] +\Lambda^m (\xi, \eta) \partial_n Y^n $ 
\raisebox{-1ex}{\rule{0cm}{4ex}}\\
 & \hspace{2.0in}$+(\bar\eta U^m) +(\bar\xi V^m)$ 
\raisebox{-2ex}{\rule{0cm}{4ex}}\\
\hline $U^m_i$ & $\delta^{q,s} (\xi, \eta) U^m (\psi) =  \partial_n \left[ U^m (\psi) \Lambda^n (\xi, \eta) - U^n (\psi) \Lambda^m (\xi, \eta)\right] +\Lambda^m (\xi, \eta) \partial_n U^n (\psi)$ 
\raisebox{-1ex}{\rule{0cm}{4ex}}\\
 & \hspace{1.0in}$+2i  (\bar\eta\gamma^a \psi) N^m_{~~a} -3i (\bar\eta\psi) R^m  
+2i  (\bar\xi\gamma^a \psi) M^m_{~~a} -2 (\bar\xi\psi) D^m $ 
\raisebox{-2ex}{\rule{0cm}{4ex}}\\
\hline $V^m_i$ & $\delta^{q,s} (\xi, \eta) V^m (\psi) =  \partial_n \left[ V^m (\psi) \Lambda^n (\xi, \eta) - V^n (\psi) \Lambda^m (\xi, \eta)\right] +\Lambda^m (\xi, \eta) \partial_n V^n (\psi) $ 
\raisebox{-1ex}{\rule{0cm}{4ex}}\\
 & \hspace{1.0in}$+2i  (\bar\xi\gamma^a \psi) N^m_{~~a} -3i (\bar\xi\psi) R^m 
-2i  (\bar\eta\gamma^a \psi) M^m_{~~a} +2 (\bar\eta\psi) D^m$ 
\raisebox{-2ex}{\rule{0cm}{4ex}}\\ 
\hline
\end{tabular}}
\newpage
\begin{center}
{\bf Table 4: Current (Non-)Conservation}
\end{center}
{\hspace{-0.1in}
\begin{tabular}{| c | l |}
\hline {\bf Noether Current} & {\hspace{1.0in}\bf Current (Non-)Conservation}    
\raisebox{-1ex}{\rule{0cm}{4ex}}\\
$J^m$ & $\partial_m J^m = \left[ \Delta {\cal L} + (\partial_m \delta x^m){\cal L}\right] -w\Gamma $
\raisebox{-2ex}{\rule{0cm}{5ex}}\\
\hline\hline $T^{m}_{~~a}$ & $\partial_m T^m_{~~a} = w^p_a (x) \Gamma$ 
\raisebox{-2ex}{\rule{0cm}{5ex}}\\
\hline $Z^m$ & $\partial_m Z^m =-w^Z (x) \Gamma$ 
\raisebox{-2ex}{\rule{0cm}{5ex}}\\
\hline $M^m_{~ab}$ & $\partial_m M^m_{~ab} (x) = -w^M_{ab} (x) \Gamma$ 
\raisebox{-2ex}{\rule{0cm}{5ex}}\\
\hline $N^m_{~~a}$ & $\partial_m N^m_{~~a} (x) = -w^n_a (x) \Gamma$  
\raisebox{-2ex}{\rule{0cm}{5ex}}\\
\hline $Q^m_i$ & $\partial_m Q^m_i = - w^{q}_i (x) \Gamma ,$ 
\raisebox{-2ex}{\rule{0cm}{5ex}}\\
\hline $S^m_i$ & $\partial_m S^m_i = - w^{s}_i (x) \Gamma ,$ 
\raisebox{-2ex}{\rule{0cm}{5ex}}\\
\hline $R^m$ & $\partial_m R^m = -w^R (x) \Gamma$ 
\raisebox{-2ex}{\rule{0cm}{5ex}}\\
\hline $D^m$ & $\partial_m D^m = -T^m_{~~a} \hat{e}_m^{~a} + Z^m \hat{e}_m^{~a} \hat\nabla_a \phi - w^D \Gamma $
\raisebox{-1ex}{\rule{0cm}{4ex}}\\
 & $= 3 \sigma \frac{\partial}{\partial \sigma} {\cal L} -w^D \Gamma $
\raisebox{-2ex}{\rule{0cm}{5ex}}\\
\hline $K^m_{~~a}$ & $\partial_m K^m_{~~a} =  6\epsilon^m x_m {\cal L} $
\raisebox{-1ex}{\rule{0cm}{4ex}}\\
 & $-T_{~~a}^m \left( 2\epsilon^a \phi [\hat{e}_m^{~b} \hat\nabla_b \phi ] + 2(\epsilon_b x^a - \epsilon^a x_b) \hat{e}_m^{~b}    -\epsilon^a \partial_m [(\bar\theta\theta) (\bar\lambda\lambda)] \right.$
\raisebox{-1ex}{\rule{0cm}{4ex}}\\
 & $\left. \qquad\qquad +(\bar\theta\theta + \bar\lambda\lambda)\epsilon^{a}_{~bc} \epsilon^c [ \hat{e}_m^{~b} -2\delta^b_{~m}] +4(\bar\theta \gamma^a \lambda)\partial_m (\bar\theta \rlap{/}{\epsilon} \lambda) \right.$
\raisebox{-1ex}{\rule{0cm}{4ex}}\\
 & $\left. \qquad\qquad\qquad\qquad - 2i \epsilon^{abc} \epsilon_c (\bar\theta \gamma_b \lambda) [\hat{e}_m^{~d}\hat\nabla_d \phi]\right)$
\raisebox{-1ex}{\rule{0cm}{4ex}}\\
 & $+ Z^m \left( 2\epsilon_a \phi \hat{e}_m^{~a} - 2i \epsilon_{mab} \epsilon^b (\bar\theta \gamma^a \lambda)
-2i (\bar\theta\theta - \bar\lambda\lambda ) [\bar\theta \rlap{/}{\epsilon} \stackrel{\leftrightarrow}{\partial}_m \lambda] \right) -w^k_{a} \Gamma$
\raisebox{-2ex}{\rule{0cm}{4ex}}\\
\hline $Y^m$ & $\partial_m Y^m = -6\phi {\cal L} +T^m_{~~a} \left( 2x^a (\hat{e}_m^{~b} \hat\nabla_b \phi) -2i \epsilon_m^{~~ab}(\bar\theta \gamma_b \lambda)\right.$
\raisebox{-1ex}{\rule{0cm}{4ex}}\\
 & $\left. \qquad\qquad +2i(\bar\theta \theta - \bar\lambda \lambda)(\bar\theta \gamma^a \stackrel{\leftrightarrow}{\partial}_m \lambda)\right) -2 Z^m \hat{e}_m^{~a} x_a  - w^Y \Gamma $ 
\raisebox{-2ex}{\rule{0cm}{4ex}}\\
\hline $U^m_i$ & $\partial_m U^m_i = \left[\Delta^u_i {\cal L} + (\partial_m \delta^u_i x^m ){\cal L}\right] - w^u_i \Gamma $\hspace{0.5in}(See Text)
\raisebox{-2ex}{\rule{0cm}{5ex}}\\
\hline $V^m_i$ & $\partial_m V^m_i = \left[\Delta^v_i {\cal L} + (\partial_m \delta^v_i x^m ){\cal L}\right] - w^v_i \Gamma $\hspace{0.5in}(See Text)
\raisebox{-2ex}{\rule{0cm}{5ex}}\\
\hline
\end{tabular}}
\medskip
~\\
\noindent
The work of MN was supported by the Japan Society for the Promotion of Science under the Post-Doctoral Research Program while that of TEC was supported in part by the U.S. Department of Energy under grant DE-FG02-91ER40681 (Task B).
\newpage

\setcounter{newapp}{1}
\setcounter{equation}{0}
\renewcommand{\theequation}{\thenewapp.\arabic{equation}}

\section*{\large\bf Appendix A: \,  Centrally Extended Superconformal Algebra}

The N=1, D=4 space-time translation generator $P^\mu$, $\mu = 0,1,2,3$,
consists of a D=3 Lorentz group vector, $p^m = P^m$, with $m=0,1,2$, and
a D=3 scalar central charge, $Z\equiv P_3$.  Likewise, the
Lorentz transformation charges $M^{\mu\nu}$ consist of two D=3 vector
representation charges: $M^{mn}=\epsilon^{mnr} M_r$ and $N^m \equiv M^{m3}$.
The $R$ charge is a singlet from both points of view.  Finally the D=4 
SUSY $(\frac{1}{2}, 0)$ spinor $Q_\alpha$ and the $(0,\frac{1}{2})$ spinor
$\bar{Q}^{\dot\alpha}$ consist of two D=3 two-component Majorana spinors:
$q_i$ and $s_i$, with $i=1,2$.  These are the charges that comprise the (centrally extended) N=2, D=3 SUSY algebra.  The spinor charges are given as linear combinations of $Q_\alpha$ and $\bar{Q}^{\dot\alpha}$ according to
\be
\pmatrix{
q_i \cr
s_i \cr} = \frac{1}{2} e^{i\frac{\pi}{4}} \pmatrix{
\sigma & i\sigma\sigma_z \cr
-i\sigma & -\sigma\sigma_z \cr} \pmatrix{Q_\alpha \cr
\bar{Q}^{\dot\alpha} \cr} ,
\ee
where $(\sigma_x, \sigma_y, \sigma_z)$ are the Pauli matrices and 
\be
\sigma \equiv \pmatrix{
1 & -1 \cr
-i & -i \cr}.
\ee
The N=1, D=4 super-Poincar\'{e} algebra can be written in terms of 
the D=3  Lorentz representation charges of the centrally extended N=2, D=3 SUSY algebra as
\begin{center}
\begin{tabular}{ll}
$[p^m , p^n ] = 0$ & $[M^m ,M^n ] = -i\epsilon^{mnr} M_r$ \\
$[p^m , Z ] = 0$ & $[M^m ,N^n ] = -i\epsilon^{mnr} N_r $ \\
& $[N^m , N^n ] = +i\epsilon^{mnr} M_r$ \\
$[M^{m}, p^n ] = -i\epsilon^{mnr} p_r$ & $[N^m , p^n ] = +i\eta^{mn} Z$ \\
$[M^{m} , Z ] = 0$ & $[N^m , Z ] = + i p^m$\\
$[M^{mn} , q_i ] = -\frac{1}{2} \gamma^{mn}_{ij} q_j\;\;\;\;\;\;\;\;\;$ &
$[N^m , q_i ] = +\frac{1}{2} \gamma^m_{ij} s_j$\\
$[M^{mn} , s_i ] = -\frac{1}{2} \gamma^{mn}_{ij} s_j$  &
$[N^m , s_i ] = -\frac{1}{2} \gamma^m_{ij} q_j$\\
$[R , q_i ] = +i s_i$ &
$\{ q_i , q_j \} = +2\left(\gamma^m C \right)_{ij} p_m$ \\
$[R , s_i ] = -i q_i$ &
$\{ s_i , s_j \} = +2\left(\gamma^m C \right)_{ij} p_m$\\
& $\{ q_i , s_j \} = -2i C_{ij} Z.$ \\
\end{tabular}
\end{center}
\be
\label{N2D3SUSY}
\ee
Note, the notation used in this paper is that of reference \cite{Clark:2002bh}, 
in particular the charge conjugation matrix and the $2+1$ (D=3) dimensional 
gamma matrices in the appropriate associated representation are given there.

The N=1, D=4 superconformal (SU$(2,2|1)$) algebra \cite{WZ}\cite{F} includes the additional charges:  
the dilatation charge $D$, a scalar from both points of view, 
the special conformal charge $K^\mu$ which
consists of a D=3 Lorentz group vector, $k^m = K^m$, with $m=0,1,2$, and
a D=3 scalar conformal central charge, $Y\equiv K_3$.  The D=4 conformal
SUSY $(\frac{1}{2}, 0)$ spinor $S_\alpha$ and the $(0,\frac{1}{2})$ spinor
$\bar{S}^{\dot\alpha}$ consist of two D=3 two-component Majorana spinors:
$u_i$ and $v_i$, with $i=1,2$.  These are the additional charges that comprise the centrally extended N=2, D=3 superconformal algebra.  The spinor charges are given as linear combinations of $S_\alpha$ and 
$\bar{S}^{\dot\alpha}$ according to
\be
\pmatrix{
u_i \cr
v_i \cr} = \frac{1}{2} e^{i\frac{\pi}{4}} \pmatrix{
\sigma & i\sigma\sigma_z \cr
-i\sigma & -\sigma\sigma_z \cr} \pmatrix{S_\alpha \cr
\bar{S}^{\dot\alpha} \cr} .
\label{confcharges}
\ee
The remaining nonzero (anti-)commutators for the N=2, D=3 centrally extended superconformal algebra are
\begin{center}
\begin{tabular}{ll}
$[M^{m}, k^n ] = -i\epsilon^{mnr} k_r\;\;\;\;\;$ & $[N^m , k^n ] = 
+i\eta^{mn} Y$ \\
$[M^{m} , Y ] = 0$ & $[N^m , Y ] = + i k^m$ \\
 & \\
$[M^{mn} , u_i ] = -\frac{1}{2} \gamma^{mn}_{ij} u_j\;\;\;\;\;\;\;\;\;$ &
$[N^m , u_i ] = +\frac{1}{2} \gamma^m_{ij} v_j$\\
$[M^{mn} , v_i ] = -\frac{1}{2} \gamma^{mn}_{ij} v_j$  &
$[N^m , v_i ] = -\frac{1}{2} \gamma^m_{ij} u_j$\\
& \\
$[R , u_i ] = -i v_i$ &
$\{ u_i , u_j \} = +2\left(\gamma^m C \right)_{ij} k_m$ \\
$[R , v_i ] = +i u_i$ &
$\{ v_i , v_j \} = +2\left(\gamma^m C \right)_{ij} k_m$\\
& $\{ u_i , v_j \} = -2i C_{ij} Y.$ \\
\end{tabular}
\end{center}
\begin{center}
\begin{tabular}{ll}
$\{ q_i , u_j \} = +2\left(\gamma^m C \right)_{ij} M_m + 2i\, C_{ij}\, D$ \\
$\{ q_i , v_j \} = +2\left(\gamma^m C \right)_{ij} N_m + 3i\, C_{ij}\, R$ \\
$\{ s_i , u_j \} = +2\left(\gamma^m C \right)_{ij} N_m + 3i\, C_{ij}\, R$ \\
$\{ s_i , v_j \} = -2\left(\gamma^m C \right)_{ij} M_m - 2i\, C_{ij}\, D$ \\
 & \\
$ [k^m, q_i]= \gamma^m_{ij} u_j $ & $[p^m, u_i]= \gamma^m_{ij} q_j $ \\
$ [k^m, s_i]= -\gamma^m_{ij} v_j $ & $[p^m, v_i]= -\gamma^m_{ij} s_j $ \\
$ [Y,q_i]=-i v_i $ & $[Z,u_i]=-i s_i $ \\
$ [Y,s_i]=-i u_i $ & $[Z,v_i]=-i q_i $ \\
 & \\
$ [D, q_i]= -\frac{i}{2} q_i $ & $[D, u_i]= +\frac{i}{2} u_i $ \\
$ [D, s_i]= -\frac{i}{2} s_i $ & $[D, v_i]= +\frac{i}{2} v_i $ \\
 & \\
$ [D, p^m]= -i p^m $ &  $[D, k^m]= +i k^m $ \\
$ [D, Z]= -i Z $ & $[D, Y]= +i Y $ \\
 & \\
$ [p^m, k^n]= 2i \left(\eta^{mn} D -\epsilon^{mnr} M_r\right) $ &  
$[Z, Y]= -2i D $ \\
$ [p^m, Y]= +2i N^m $ & $[k^m,Z]= +2i N^m $
\end{tabular}
\end{center}
\be
\label{N2D3Superconformal}
\ee
\newpage

\setcounter{newapp}{2}
\setcounter{equation}{0}
\renewcommand{\theequation}{\thenewapp.\arabic{equation}}

\section*{\large\bf Appendix B: \, Superconformal Transformations}

The presence of the non-BPS brane in N=1, D=4 superspace spontaneously breaks the super-Poincar\'e group to the D=3 Poincar\'e symmetry and $R$ symmetry groups.  Commensurate with this is the appearance of the broken target space translation symmetry Nambu-Goldstone boson $\phi$ and the broken SUSY Goldstino D=3 Majorana spinor fields $\theta_i$ and $\lambda_i$, $i=1,2$.  Recalling the coset method of construction for the realization of this spontaneously broken super-Poincar\'e group on these fields from reference \cite{Clark:2002bh}, the world volume coordinate variation and the linearly represented total variations of the fields were found to be 
\bea
x^{\prime m} &=& x^m + a^m  - i (\bar\xi \gamma^m \theta + \bar\eta \gamma^m 
\lambda ) -\phi b^m + \epsilon^{mnr} \alpha_n x_r \cr
\phi^\prime &=& \phi + z + (\xi \gamma^0 \lambda - \theta \gamma^0 \eta ) - b^m 
x_m \cr
\theta^\prime_i &=& \theta_i+\xi_i - \frac{i}{2} b_m (\gamma^m \lambda )_i - 
i\rho 
\lambda_i  -\frac{i}{2} \alpha_m (\gamma^m \theta )_i \cr
\lambda^\prime_i &=& \lambda_i+\eta_i + \frac{i}{2} b_m (\gamma^m \theta )_i + 
i\rho 
\theta_i  -\frac{i}{2} \alpha_m (\gamma^m \lambda )_i  .
\label{transN2D3}
\eea
Thus, applying the relation between total and intrinsic variations of a field, $\delta \varphi = \Delta \varphi - \delta x^m \partial_m \varphi$, the nonlinear realization of the N=2, D=3 super-Poincar\'{e} algebra, given in equation (\ref{N2D3SUSY}), on the $\phi$, $\theta$ and $\lambda$ fields is
\begin{center}
\begin{tabular}{ll}
$\delta^p(a) \phi = -a^m \partial_m \phi $ &\quad $\delta^Z(z) \phi = z$ \\
$\delta^p(a) \theta =  -a^m \partial_m \theta $ &\quad $\delta^Z(z) \theta = 0$ 
\\
$\delta^p(a) \lambda = -a^m \partial_m \lambda $  &\quad $\delta^Z(z) \lambda = 0$ 
\\
 & \\
$\delta^q(\xi) \phi = -\bar\xi \lambda +i\bar\xi \gamma^m \theta 
\partial_m \phi $  &\quad $\delta^s(\eta) \phi = +\bar\eta \theta +i\bar\eta 
\gamma^m \lambda \partial_m \phi $  \\
$\delta^q(\xi) \theta_i = \xi_i +i\bar\xi \gamma^m \theta \partial_m 
\theta_i $ &\quad $\delta^s(\eta) \theta_i = +i\bar\eta \gamma^m \lambda 
\partial_m \theta_i $ \\
$\delta^q(\xi) \lambda_i = +i\bar\xi \gamma^m \theta \partial_m \lambda_i $ &\quad $\delta^s(\eta) \lambda_i = \eta_i +i\bar\eta \gamma^m \lambda \partial_m 
\lambda_i $  \\
\end{tabular}
\end{center}
\begin{center}
\begin{tabular}{ll}
$\delta^M(\alpha) \phi = -\epsilon^{mnr} \alpha_n x_r \partial_m \phi $ &\quad 
$\delta^N(b) \phi = -b^m x_m  +\phi b^m \partial_m \phi $  \\
$\delta^M(\alpha) \theta_i = -\frac{i}{2}\alpha_m (\gamma^m \theta)_i -
\epsilon^{mnr} \alpha_n x_r \partial_m \theta_i $  &\quad $\delta^N(b) \theta_i = 
+\frac{i}{2}b_m (\gamma^m \lambda)_i +\phi b^m 
\partial_m \theta_i $  \\
$\delta^M(\alpha) \lambda_i = -\frac{i}{2}\alpha_m (\gamma^m \lambda)_i -
\epsilon^{mnr} \alpha_n x_r \partial_m  \lambda_i $  &\quad $\delta^N(b) \lambda_i 
= -\frac{i}{2}b_m (\gamma^m \theta)_i +\phi b^m 
\partial_m \lambda_i $ \\
 & \\
$\delta^R(\rho) \phi = 0$ & \\
$\delta^R(\rho) \theta_i = -i\rho \lambda_i$ & \\
$\delta^R(\rho) \lambda_i = +i\rho \theta_i$ . & \\
\end{tabular}
\end{center}
\be
~
\ee

The brane tension explicitly breaks the superconformal symmetries at low energies (not to mention the hard superconformal symmetry breaking by radiative corrections at all scales).  Hence, no new Nambu-Goldstone fields are expected to arise from the spontaneously broken component of these hard broken superconformal symmetries.  Yet, in addition to the explicit hard breaking, the $Y$, $u_i$ and $v_i$ conformal symmetries have a spontaneously broken component as required by the superconformal algebra.  Consider the ground state, $|0>$, expectation value of the Jacobi identity involving $p^m$, $Y$ and $\partial^n \phi$ along with the fact that the brane is D=3 translation invariant, $p^m |0>=0$, and the superconformal algebra contains the commutator $[p^m , Y]=2iN^m$ while the D=3 translations are represented by world volume derivatives, $[p^m ,\partial^n \phi] = i\partial^m \partial^n \phi$, to obtain
\be
<0|[Y, \partial^m \partial^n \phi] |0> = -2 <0|[N^m, \partial^n \phi] |0>.
\label{YGB}
\ee
On the other hand, the Jacobi identity involving $p^n$, $N^m$ and $\phi$ along with the algebraic relation $[N^m , p^n] =i\eta^{mn}Z$ implies that
\be
<0| [N^m , i\partial_n \phi] |0> = -i\delta^{m}_{~n}<0| [Z, \phi] |0> = \delta^{m}_{~n} \neq 0,
\ee
the right hand side being nonzero since $\phi$ is the $Z$ symmetry Nambu-Goldstone boson which transforms into a constant.  Hence, the $N^m$ symmetry has a spontaneously broken component with $\partial^m \phi$ acting as the corresponding Nambu-Goldstone boson.  Further, from equation (\ref{YGB}), it is found that the $Y$ symmetry is spontaneously broken with $\partial^2 \phi$ acting as the corresponding Nambu-Goldstone boson.  Likewise, the conformal SUSY symmetries are spontaneously broken with derivatives of $\theta$ and $\lambda$ acting as their Goldstino modes.  Even though the conformal central charge symmetry $Y$, the broken Lorentz symmetry generator $N^m$ as well as the conformal SUSY charges $u$ and $v$ all have a spontaneously broken symmetry component, no new Nambu-Goldstone modes are required to realize these spontaneously broken symmetry transformations.

Thus the remaining superconformal transformations are to be realized on the $\phi$, $\theta$ and $\lambda$ Nambu-Goldstone fields.  These transformations can be most easily found by using the coset construction with a considerable simplification.  Consider the coset element $\Omega \in G/H$ with $G= SU(2,2|1)$, the superconformal group, and $H$ generated by the set of unbroken charges $\{M^{mn}, R, D, k^m\}$.  $\Omega$ can be written as the product $\Omega=\Omega_o \hat\Omega$ where
\be
\Omega_o = e^{ix^mp_m} e^{i(\phi Z+\bar\theta q + \bar\lambda s)} = e^{ix^\mu P_\mu} e^{i(\theta Q +\bar\theta \bar{Q})}
\ee
and 
\be
\hat\Omega = e^{iv^m N_m} e^{i( fY +\bar\chi u +\bar\psi v)}.
\ee
$g\in G$ acts on $\Omega$ to yield $g\Omega =\Omega_o^\prime \hat{h}\hat\Omega = \Omega_o^\prime \hat\Omega^\prime h$ with 
$h\in H$ while $\hat{h}$ is generated by $\{M^{mn}, N^m, R, D, k^m, Y, u, v\}$.  The simplification in finding the superconformal transformations of $\phi$, $\theta$ and $\lambda$ comes from the observation
\be
\hat{h} \hat\Omega = \hat\Omega^\prime h ~~;
\ee
that is $\hat{h}$ acting on $\hat\Omega$ does not feedback to a change in $\Omega_o^\prime$.  Hence, as far as the superconformal transformations of $\phi$, $\theta$ and $\lambda$ are concerned, only the action of $g\Omega_o$ needs to be considered.  With $g$ a superconformal transformation generated by $\{D, K^\mu, S^\alpha, \bar{S}_{\dot\alpha}\}$ the transformation has the form
\be
g \Omega_o = e^{i x^{\prime \mu} P_\mu} e^{i(\theta^\prime Q + \bar\theta^\prime \bar{Q})} \hat{h} .
\ee
This is just the usual linear representation of a superconformal motion in N=1, D=4 superspace.  For the transformations
\be
g= e^{i(\epsilon D + \epsilon^\mu K_\mu + \eta S + \bar\eta \bar{S})}
\ee
the superspace motion is given by \cite{Clark:1978jx}\cite{PS1}
\bea
x^{\prime \mu} &=& (1+\epsilon)x^\mu +2\epsilon^\rho x_\rho x^\mu - x^2 \epsilon^\mu +\theta^2 \bar\theta^2 \epsilon^\mu \cr
 & & -ix^\mu \eta\theta - x_\rho (\theta\sigma^{\mu\rho} \eta ) -\theta^2 (\eta \sigma^\mu \bar\theta ) \cr
 & & +ix^\mu \bar\eta \bar\theta + x_\rho (\bar\eta \bar\sigma^{\rho\mu}\bar\theta ) + \bar\theta^2 (\bar\eta \bar\sigma^\mu \theta ) \cr
\theta^{\prime \beta} &=& (1+\frac{1}{2} \epsilon ) \theta^\beta + \epsilon^\rho x_\rho \theta^\beta -i \epsilon^\mu x^\rho (\theta \sigma_{\mu\rho})^\beta + 2i (\theta \rlap{/}{\epsilon} \bar\theta)\theta^\beta \cr
 & & -2i\theta^2 \eta^\beta -(\rlap{/}{x} \bar\eta)^\beta +2i (\bar\eta \bar\theta) \theta^\beta \cr
\bar\theta^{\prime \dot\beta} &=& (1+\frac{1}{2} \epsilon ) \bar\theta^{\dot\beta} + \epsilon^\rho x_\rho \bar\theta^{\dot\beta} -i \epsilon^\mu x^\rho (\bar\theta \bar\sigma_{\mu\rho})^{\dot\beta} - 2i (\theta \rlap{/}{\epsilon} \bar\theta)\bar\theta^{\dot\beta} \cr
 & & +2i\bar\theta^2 \bar\eta^{\dot\beta} -(\eta \rlap{/}{x})^{\dot\beta} - 2i (\eta \theta) \bar\theta^{\dot\beta} .
\label{sconftrans}
\eea

Recalling that $x^3 = \phi$, $P_3 =Z$, $K_3 =Y$, etc., while the N=1, D=4 superspace Grassmann coordinates $\theta^\alpha$ and $\bar\theta_{\dot\alpha}$ are related to the N=2, D=3 Goldstinos $\theta_i$ and $\lambda_i$ according to 
\be
\pmatrix{
\theta^\alpha \cr
\bar\theta_{\dot\alpha} \cr} = \frac{1}{2} e^{i\frac{\pi}{4}} \pmatrix{
\sigma_z \sigma^{\rm T} (\theta -i \lambda) \cr
i\sigma^{\rm T} (\theta + i \lambda) \cr} ,
\ee
and the conformal SUSY charges obey the inverse relation to equation (\ref{confcharges})
\be
\pmatrix{
S_\alpha \cr
\bar{S}^{\dot\alpha} \cr} = \frac{1}{2} e^{i\frac{\pi}{4}} \pmatrix{
-i\sigma^{\rm T} \sigma_z (u + iv) \cr
-i\sigma^{\rm T} \sigma_x (u - iv) \cr} ,
\ee
the various N=2, D=3 superconformal variations of the Nambu-Goldstone fields are determined.  Isolating the dilatations with parameter $\epsilon$, the world volume coordinate transformation and total variation of the fields are found
\bea
\delta^D (\epsilon) x^m &=& \epsilon x^m \cr
\Delta^D (\epsilon) \phi &=& \epsilon \phi \cr
\Delta^D (\epsilon) \theta_i &=& \frac{1}{2} \epsilon \theta_i \cr
\Delta^D (\epsilon) \lambda_i &=& \frac{1}{2} \epsilon \lambda_i .
\eea
In general the (Lie) intrisic variation of a field $\varphi$ is related to the total variation of the field according to $\delta \varphi = \Delta \varphi - \delta x^m \partial_m \varphi$.  Hence the intrinsic dilatation transformation has the linear representation
\bea
\delta^D (\epsilon) \phi &=& \epsilon(1-x^m \partial_m ) \phi \cr
\delta^D (\epsilon) \theta_i &=&  \epsilon ( \frac{1}{2} -x^m \partial_m ) \theta_i \cr
\delta^D (\epsilon) \lambda_i &=& \epsilon ( \frac{1}{2} -x^m \partial_m ) \lambda_i .
\eea

The special conformal transformations with generators $k^m$ and conformal central charge transformations with charge $Y$ are obtained from equation (\ref{sconftrans}) by means of the identification $\epsilon^\mu =(\epsilon^m ,  y)$ and $ \epsilon^\mu K_\mu = \epsilon^m k_m + yY$
\bea
\delta^Y (y) x^m &=& -2y\phi x^m \cr
\Delta^Y (y) \phi &=& -y ( x^2 + \phi^2 + (\bar\theta \theta)(\bar\lambda \lambda) ) \cr
\Delta^Y (y) \theta &=& y ( i(\rlap{/}{x} \lambda ) -\phi \theta - (\bar\theta \theta)\lambda ) \cr
\Delta^Y (y) \lambda &=& y ( - i(\rlap{/}{x} \theta ) -\phi \lambda + (\bar\lambda \lambda)\theta ) \cr
 & & \cr
\delta^k (\epsilon) x^m &=& 2 \epsilon^n x_n x^m - x^2 \epsilon^m + \phi^2 \epsilon^m - (\bar\theta \theta)(\bar\lambda \lambda) \epsilon^m \cr
\Delta^k (\epsilon) \phi &=& 2 \epsilon^n x_n \phi \cr
\Delta^k (\epsilon) \theta &=& \epsilon^n x_n \theta - 2i (\bar\theta \rlap{/}{\epsilon} \lambda ) \lambda -i \phi (\rlap{/}{\epsilon} \lambda ) -i \epsilon^{mnr} \epsilon_m x_n (\gamma_r \theta ) \cr
\Delta^k (\epsilon) \lambda &=& \epsilon^n x_n \lambda + 2i (\bar\theta \rlap{/}{\epsilon} \lambda ) \theta +i \phi (\rlap{/}{\epsilon} \theta ) - i \epsilon^{mnr} \epsilon_m x_n (\gamma_r \lambda ) .
\eea

Finally the conformal SUSY transformations with charges $u_i$ and $v_i$ are determined to be, using explicit indices, 
\bea
\delta^u_i x^m &=& - x^m \theta_i + i (\bar\lambda \lambda) (\gamma^m \theta)_i + i\phi (\gamma^m \lambda)_i + i\epsilon^{mnr} x_n (\gamma_r \theta)_i \cr
\Delta^u_i \phi &=& +i (\rlap{/}{x} \lambda)_i -\phi \theta_i -(\bar\theta \theta ) \lambda_i \cr
\Delta^u_i \theta_j &=& + i(\rlap{/}{x} C)_{ij} +\frac{1}{2} (\bar\theta \theta) C_{ij} -\frac{3}{2} (\bar\lambda \lambda) C_{ij} \cr
\Delta^u_i \lambda_j &=& + \phi C_{ij} + 2 (\bar\theta \lambda ) C_{ij} - [\theta_i \lambda_j - \lambda_i \theta_j ] \cr
 & & \cr
\delta^v_i x^m &=& + x^m \lambda_i - i (\bar\theta \theta) (\gamma^m \lambda)_i + i\phi (\gamma^m \theta)_i - i\epsilon^{mnr} x_n (\gamma_r \lambda)_i \cr
\Delta^v_i \phi &=& +i (\rlap{/}{x} \theta)_i +\phi \lambda_i -(\bar\lambda \lambda ) \theta_i \cr
\Delta^v_i \theta_j &=& + \phi C_{ij} - 2 (\bar\theta \lambda ) C_{ij} - [\theta_i \lambda_j - \lambda_i \theta_j ] \cr
\Delta^v_i \lambda_j &=& - i(\rlap{/}{x} C)_{ij} +\frac{3}{2} (\bar\theta \theta) C_{ij} -\frac{1}{2} (\bar\lambda \lambda) C_{ij} .
\eea

As usual, the intrinsic variations can be found according to $\delta \varphi = \Delta \varphi - \delta x^m \partial_m \varphi$ for each field $\varphi$.  Note that $Y$, $k^m$, $u_i$ and $v_i$ variations are all nonlinearly realized, however the special conformal symmetry generated by $k^m$ is not spontaneously broken while the $Y$, $u_i$ and $v_i$ generated symmetries are spontaneously broken due to the field independent but world volume coordinate dependent terms in their transformation equations as discussed above.

\pagebreak

\newpage
\end{document}